\documentclass[preprint]{vgtc}               

\ifpdf
  \pdfoutput=1\relax                   
  \usepackage{graphicx}                
  \DeclareGraphicsExtensions{.pdf,.png,.jpg,.jpeg} 
\else
  \usepackage{graphicx}   
  \DeclareGraphicsExtensions{.eps}     
\fi%

\graphicspath{{figures/}{pictures/}{images/}{./}} 

\usepackage{microtype}                 
\PassOptionsToPackage{warn}{textcomp}  
\usepackage{textcomp}                  
\usepackage{mathptmx}                  
\usepackage{times}                     
\usepackage{cite}                      
\usepackage{flushend}
\usepackage{tabu}                      
\usepackage{booktabs}                  
\usepackage{xcolor}

\usepackage{caption}
\usepackage{subcaption}

\onlineid{0}

\vgtccategory{Research}

\vgtcinsertpkg

\preprinttext{To be published at IEEE Xplore.}

\newcommand{\SC}{Story Creator \space}
\newcommand{\SV}{Story Viewer \space}

\title{A Workflow Approach to Visualization-Based Storytelling \\ with Cultural Heritage Data}

\author{Johannes Liem\thanks{e-mail: johannes.liem@donau-uni.ac.at}\\ %
        \parbox{1.4in}{\scriptsize \centering University for Continuing \\ Education Krems} %
\and Jakob Kusnick\thanks{e-mail: kusnick@imada.sdu.dk}\\ %
     \parbox{1.4in}{\scriptsize \centering University of \\ Southern Denmark} %
\and Samuel Beck\thanks{e-mail: samuel.beck@vis.uni-stuttgart.de}\\ %
     \scriptsize University of Stuttgart %
\and Florian Windhager\thanks{e-mail: florian.windhager@donau-uni.ac.at}\\ %
     \parbox{1.4in}{\scriptsize \centering University for Continuing \\ Education Krems} %
\and Eva Mayr\thanks{e-mail: eva.mayr@donau-uni.ac.at}\\ %
     \parbox{1.4in}{\scriptsize \centering University for Continuing \\ Education Krems}}

\abstract{Stories are as old as human history–-and a powerful means for the engaging communication of information, especially in combination with visualizations. The InTaVia project is built on this intersection and has developed a platform which supports the workflow of cultural heritage experts to create compelling visualization-based stories: From the search for relevant cultural objects and actors in a cultural knowledge graph, to the curation and visual analysis of the selected information, and to the creation of stories based on these data and visualizations, which can be shared with the interested public.%
} 

\CCScatlist{
  \CCScatTwelve{Human-centered computing}{Visu\-al\-iza\-tion}{Visu\-al\-iza\-tion appli\-ca\-tion domains}{};
  \CCScatTwelve{Applied computing}{Arts and humanities}{}}

\begin{document}


\firstsection{Introduction}

\maketitle

Their omnipresence in human culture---as well as reflections from various scholarly perspectives---make clear: Stories or narratives are among the most essential design strategies for conveying novel, relevant or entertaining information both in present-day culture and throughout human history \cite{bolin2010re,dykes2019effective}. Therefore, storytelling is also an important design strategy to communicate cultural heritage information---from the design of physical exhibitions to digital knowledge communication initiatives, and also in hybrid settings. 

For the public communication of data-rich subject matters, storytelling has become a ubiquitous topic and strategy in visualization research and development \cite{segel2010narrative, riche2018data}. There are signs for similar developments in arts, humanities, and cultural heritage (CH) fields, where historical sources and information about artists and cultural objects have been digitized and made successively available for a variety of interested user groups. 

To fully tap into the potential of these cultural heritage data for visualization-based storytelling, it is necessary to take the intertwined workflows of visual data exploration and story creation into account: Many existing tools provide the means for the actual creation of stories, but do not support preceding practices of searching for CH data, of assembling and curating topic-centered data collections, and for their (visual) analysis. However, stories about cultural heritage topics (be it as an exhibition in a museum or in the digital world) are usually the result of a research process, where curators search for relevant cultural objects and background information, analyze it in depth and finally curate it, before they translate this information and their insights into a compelling narration.

\begin{figure}
 \centering
 \includegraphics[width=\columnwidth]
{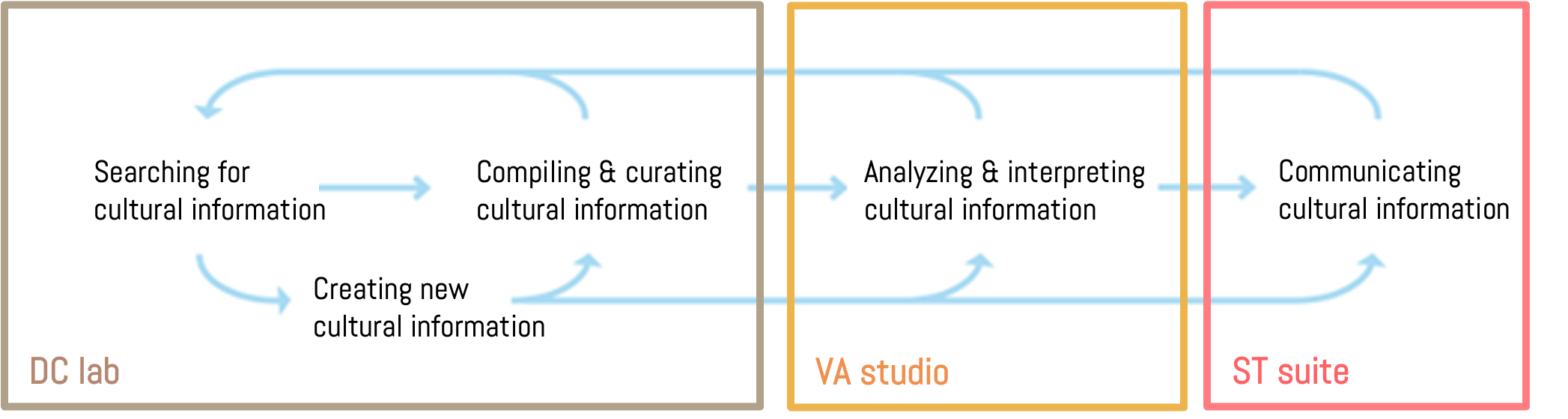}
 \caption{Iterative workflow model guiding the development of the InTaVia platform (arrows in blue), annotated with the main modules data curation lab (DC lab), visual analytics studio (VA studio) and storytelling suite (ST suite) that cover each stage of this workflow.}
 \label{fig:01}
\end{figure}

The H2020-project InTaVia (In/Tangible Cultural Heritage: Visual Analysis, Curation \& Communication) aims to support this multi-stage workflow by (i) assembling a transnational knowledge graph, which integrates information on cultural objects and cultural actors, (ii)  by creating an intuitive interface for searching information within this knowledge graph, (iii) a visual analytics studio for the visual analysis and curation of this information, and finally (iv) a visual storytelling suite for their translation into compelling visualization-based stories. 
Figure~\ref{fig:01} provides an overview of the practice and workflow model guiding the InTaVia project, to which the overall architecture of the platform responds.
The design and development of the platform was informed by ideation and evaluation workshops where we collected feedback from CH domain experts and created case studies with them (e.g., “Traveling with Albrecht Dürer”, Section~\ref{sec:caseStudyAlbrecht}).

\begin{figure*}
 \centering
 \includegraphics[width=\textwidth]
{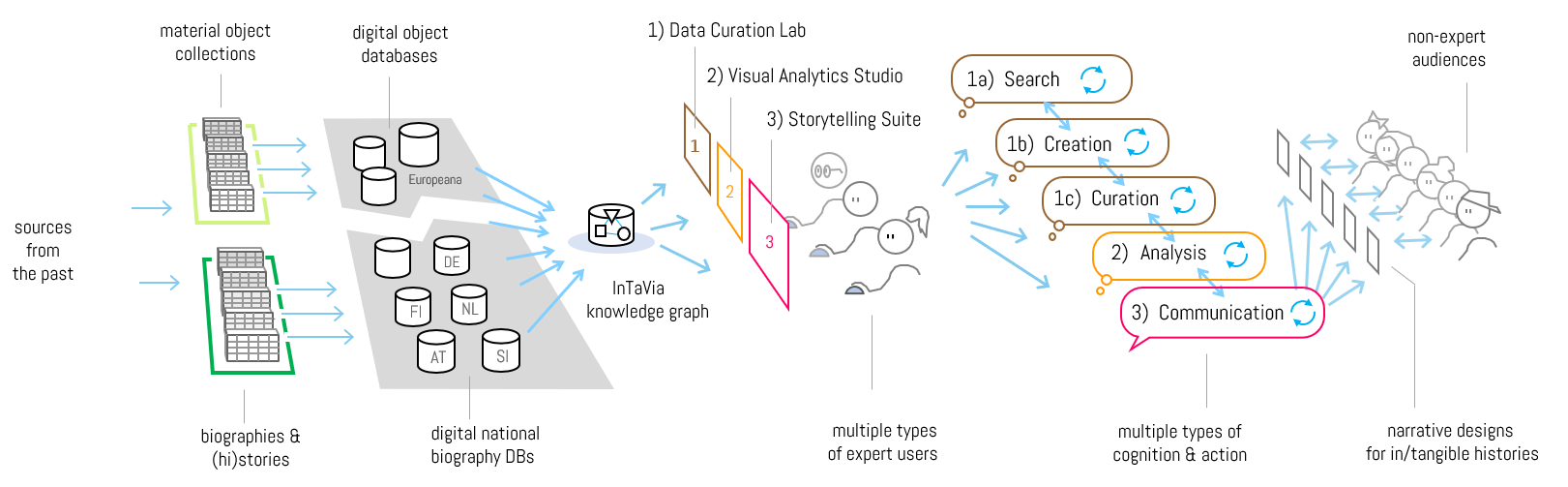}
 \caption{Architecture of and information flow within the InTaVia platform, supporting a variety of cultural heritage data practices with visualization-based interfaces, including activities of searching, creating, curating, analyzing, and communicating for a large variety of user groups.}
 \label{fig:02}
\end{figure*}


In the following, we first discuss related work on workflows in visualization-based storytelling. Then we reflect on the implementation of components supporting these workflows within the InTaVia platform as a whole, before we focus on the visualization-based storytelling components that draw together interactive visualizations and rich-media content to interweave them with narrative text annotations. Thus, InTaVia facilitates and fosters data-driven storytelling in a wide range of arts, history and humanities fields, drawing on the modular and sequential architecture of InTaVia to support the whole workflow when creating stories on cultural heritage information.

\section{Workflows in Visualization-Based Storytelling}

Considerations from different domains show that combinations of stories and visualizations are well-suited to convey relevant overviews and essential findings or details in an entertaining and memorable way \cite{bolin2010re,dykes2019effective,segel2010narrative, riche2018data}. To efficiently generate such insightful stories based on visualizations, several approaches and workflows were proposed. 

Looking at the process of generating visualization-based stories, Dykes \cite{dykes2019effective} describes the procedure as a series of successive steps, which trigger one another "like a line of dominos" (p. 33): Data has to be (i) collected and (ii) organized to (iii) gain insight, which then (iv) can be communicated in a data story to recipients. Even clearer, Lee et al. \cite{lee2015more} describe the generation of stories as a three-phased visual data storytelling process: (i) explore the data, (ii) make a story, and (iii) tell the story. 

Recent approaches in narrative digital humanities also automatize parts of these processes to automatically generate structured story points from large knowledge bases, or to extract them from texts for their subsequent visualization \cite{bartalesi2023unstructured}.
In other fields (e.g., data journalism) there are similar visualization-based storytelling systems with similar layout and workflows,  such as for the analysis and content extraction of social media data~\cite{scharl2019multimodal}.
However, aside from early, experimental work on fully AI-driven story creation, human minds remain the main drivers of the outlined multi-stage workflows.

While the corresponding process models seem to follow a linear order at first sight, the actual workflows require various iterations and cycles of re-exploring and re-collecting further data for identifying and generating a story \cite{lee2015more}. This is why a close interconnection of modules for querying, curating, and analyzing data, together with a module for the creation of visualization-based stories might suit these workflows better and has been chosen in the InTaVia project.

\begin{figure}[hb]
    \centering
    \includegraphics[width=\columnwidth]{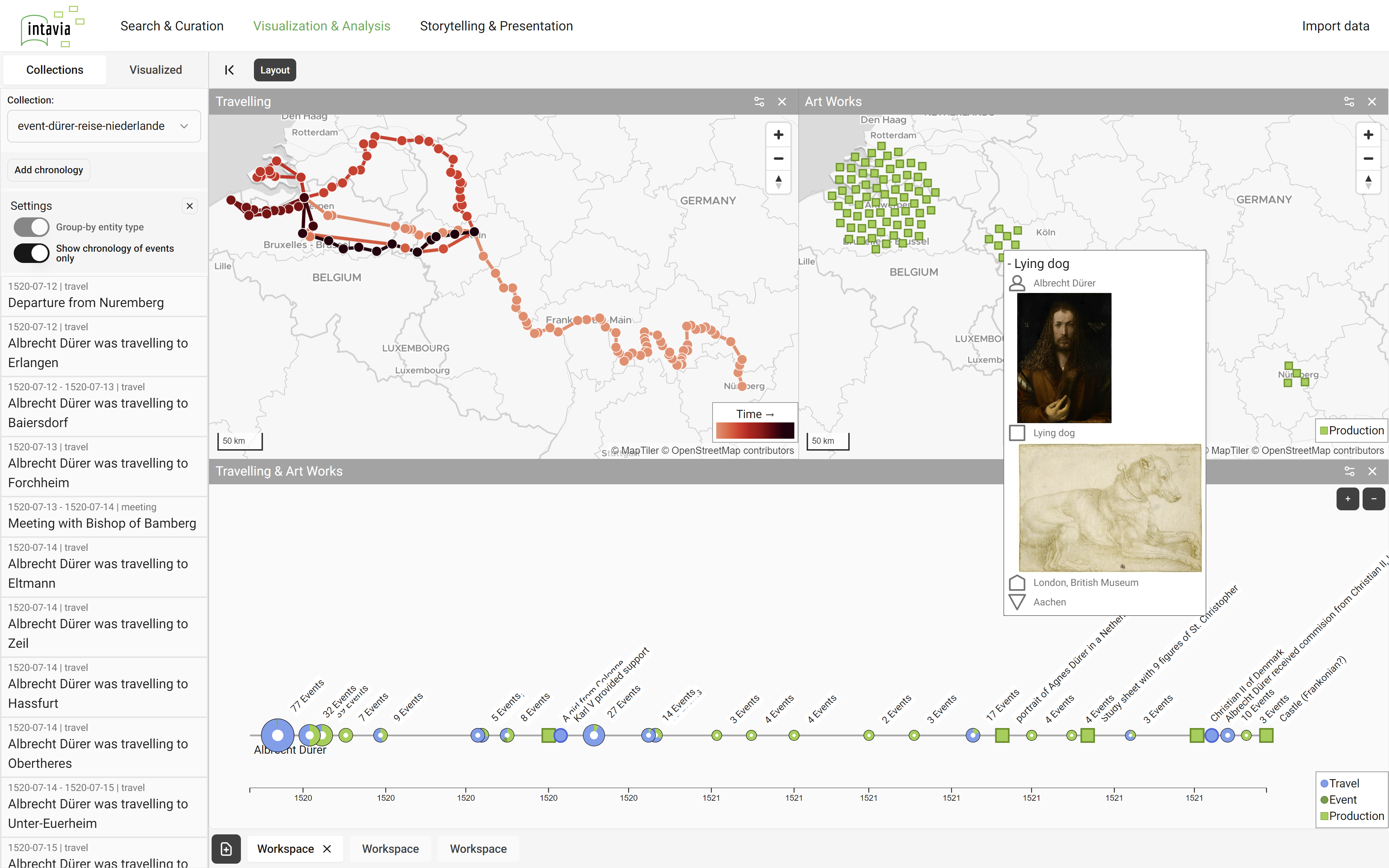}
    \caption{Dürer's journey to The Netherlands (1520-1521) in space and time including travel directions, stops and events, and produced art works along the way. Visualized in the Visual Analytics Studio. Based on manually curated data \cite{grebeNL2013}.}
    \label{fig:vas}
\end{figure}

\begin{figure*}[ht!]
    \centering
    \includegraphics[width=\textwidth]{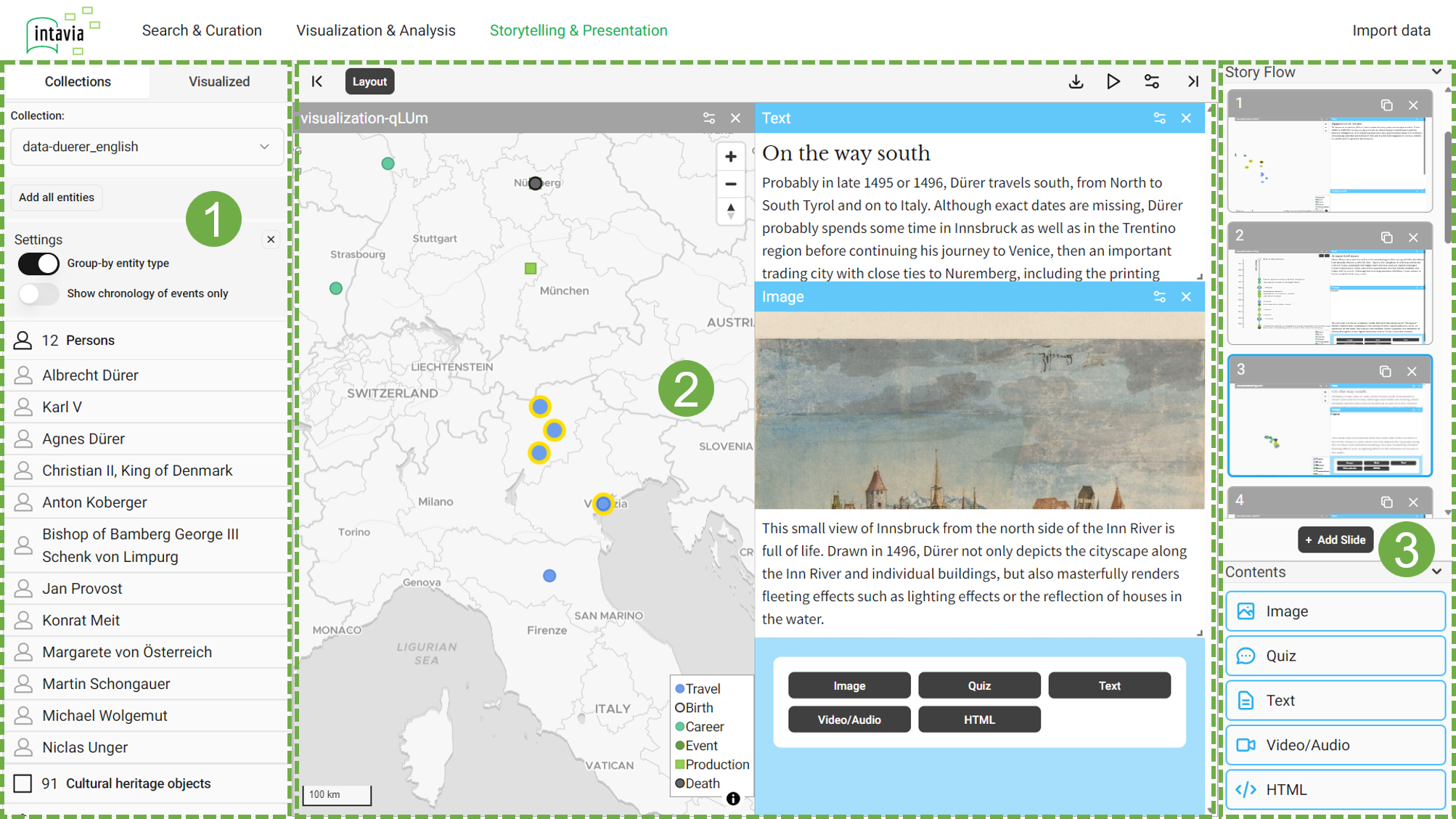}
    \caption{Overview of the \SC interface: (1) data panel containing collections, entities, and events, (2) main panel for adding visualizations and media content to slides, (3) story flow panel with slide overview and content toolbar. The selected and highlighted event dots on the map are zoomed and panned into the center of the screen during the presentation of slides in the \SV to set the focus on these specific events.}
    \label{fig:sc-overview}
\end{figure*}

\section{The InTaVia Platform: Knowledge Graph, Curation \& Visualization}

The H2020-project InTaVia (``In/Tangible European Heritage – Visual Analysis, Curation \& Communication'', \url{https://intavia.eu}) develops a platform for the visual analysis, curation and communication of CH information. As a main source, it has assembled a transnational and multimodal knowledge graph for cultural heritage data to counteract some of the structural problems resulting from siloed and separated data collections in digital cultural heritage realms \cite{mayr2022multiple,smashingsilos}. Among these problems, it primarily works to overcome the separation of databases for a) ``tangible'' cultural objects (such as paintings, sculptures, buildings or literary texts) and b) for ``intangible'', contextual information, such as the biographical information on cultural actors and artists contained in biographical and prosopographical lexica. The InTaVia knowledge graph draws together data from both types of knowledge collections and currently includes 22,347,784 triples on 111,551 actors from four different European prosopographical data sources (Austria, Finland, the Netherlands, and Slovenia) with data on 172,370 related cultural heritage objects from Wikidata and Europeana \cite{dariah}. Next to information about persons and cultural heritage objects, the knowledge graph includes entities for institutions (e.g., academies or universities), historical events (e.g., wars), and places. Whether person, assembly, or thing - InTaVia treats each of these entities as a potential protagonist of a story, so that it can have a history of ``biographical'' events (e.g., birth, creation, travel), including time stamps and relations to other entities.

The system's architecture (Figure~\ref{fig:02}) was designed to support our guiding workflow model (Figure~\ref{fig:01}) with cultural heritage data for users of the InTaVia frontend\footnote{\url{https://intavia.acdh-dev.oeaw.ac.at/}}. For the first step in the workflow, querying the data, users can constrain query parameters (e.g., names, occupations, date ranges) either by form fields or by interactive visualizations (visual query builder with ``scented widgets'' \cite{willett2007scented,ciorna2023tour}). For the creation (second step) and curation (third step) of data, users are enabled to create new data, or edit and enrich existing data sets locally, in the module of InTaVia's ``data curation lab''.

For the fourth step in the workflow, data can be visually analyzed either for individual entities from an ego-perspective in a detail view (Figure~\ref{fig:vas}) or for multiple entities in linked coordinated views \cite{windhager2022visuelle}. Different types of interactive visualizations are available based on the specific data and with regard to CH experts’ related research questions: timelines, maps, and network visualizations. 
After they explored and gained insights on the data, users can move to the fifth step of the workflow model and assemble the results of their visually supported query, curation and analysis activities into visualization-based stories.

\section{The InTaVia Storytelling Suite}

The Storytelling Suite implements a two-staged storytelling process by the means of two functional sub-modules:
The \textbf{\SC} allows users to create dynamic slideshow-based stories \cite{roth2021carto} and the \textbf{\SV} displays the resulting interactive stories in desktop and mobile browsers.

Both modules are developed as responsive JavaScript web applications using contemporary frameworks, including React \cite{react} and Vue.js \cite{vue} for the interface, and MapLibre \cite{maplibre} and D3.js \cite{d3} for the interactive visualizations.

\begin{figure*}[ht]
    \centering
    \begin{subfigure}{0.72\textwidth}
        \centering
        \includegraphics[width=\textwidth]{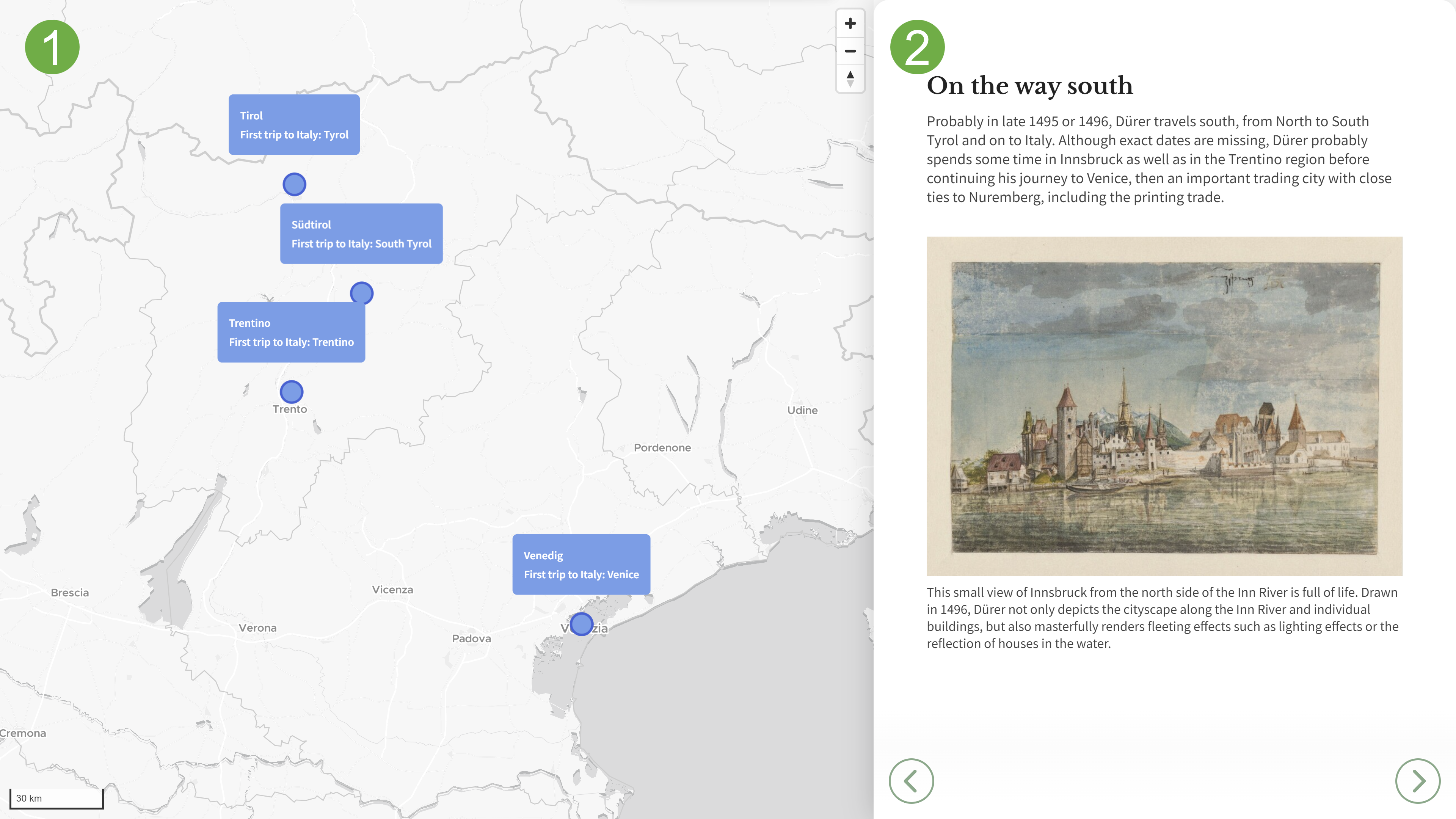}
        \caption{Desktop layout}
        \label{subfig:sv-desktop}
    \end{subfigure}
    \color{lightgray}{\vrule} \hfill 
    \begin{subfigure}{0.23\textwidth}
        \centering
        \includegraphics[width=\textwidth]{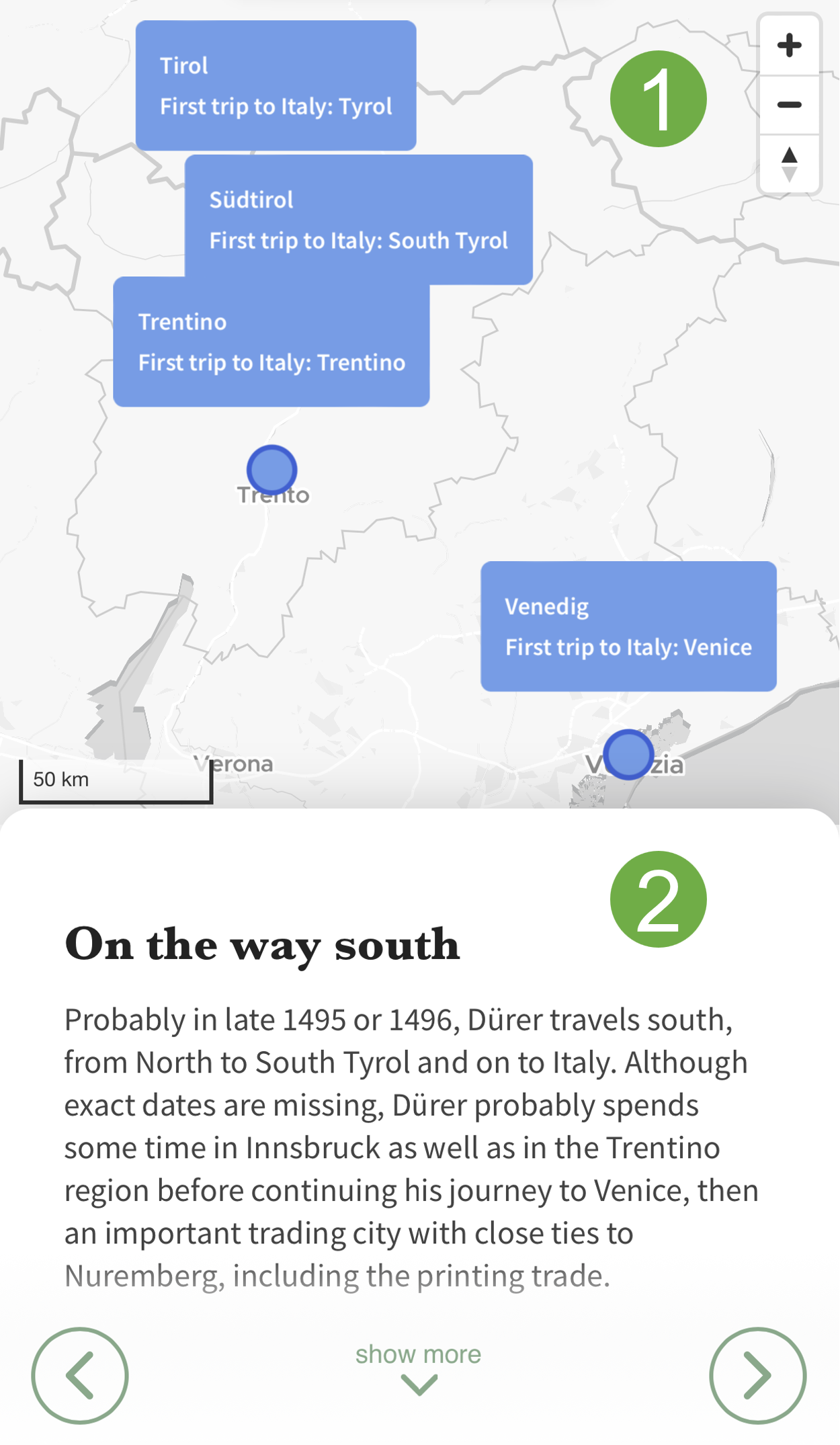}
        \caption{Mobile layout}
        \label{subfig:sv-mobile}
    \end{subfigure}
    \caption{Overview of the \SV interface on desktop (left) and mobile devices (right) which organize map (1) and media content (2) either next to each other or in an expandable panel. The four visited and selected stations on Albrecht D\"urer's travel to Italy are in the focus and annotated by texts and an image for the tangible narration.}
    \label{fig:sV}
\end{figure*}

\subsection{Story Creator}

The \SC is the authoring component of the Visual Storytelling Suite, providing a user-friendly interface for creating and editing slide-based stories enabling the seamless integration of visualizations and other multimedia elements into the story.
The data basis for stories is either originating from the InTaVia knowledge graph or from manually curated and imported data, which is then utilized in the following workflow.
Users can initiate new stories or re-import previously exported stories.
Throughout the project, we developed several multimodal, representative showcases that are available on the overview page. 
These stories provide a summary of functionalities and aim to inspire the authoring of new stories.
By clicking on a story name, users can access the \SC and make changes to the content chunks and story flow including their embedded visualizations.

\textbf{Content Creation and Editing.} At the core of the \SC is the slide editor (Figure~\ref{fig:sc-overview} (2)), where users define visualizations and content chunks within a slide. 
Users can either incorporate visualizations created specifically for the story or reuse visualizations created during the prior analysis in the visual analytics studio.
Each story slide has a predefined, yet user-selected layout dividing a slide into areas for visualizations and content chunks. 
We predefined a set of layouts which also work well on mobile devices. 
To ensure this, we limited the number of possible visualizations to one per slide, whereby a maximum number of two content panels (able to hold multiple elements) is allowed to enable the detailed discourse on multimedia content chunks.
Users can customize the layout of content elements through drag and drop interactions within a slide's grid.

\textbf{Slide Management and Flow Control.}
In the story flow panel each slide is represented by a thumbnail card, which can be duplicated, deleted by buttons or rearranged using drag and drop interactions. 
The \SC incorporates a further feature called \textit{nested slides} that allows users to create drill-down stories \cite{segel2010narrative} to optionally provide detailed narration steps between the current and next slide. 
This feature enhances the storytelling experience by enabling users to present additional information or delve into specific details on demand without interrupting the flow of the main narrative. 
Nested slides are useful for providing context, explanations, or supplementary content within a specific segment of the story. 
The inclusion of the nested slides feature empowers users to create multi-layered narratives, offering flexibility and depth in presenting information, and providing a more flexible, immersive and interactive storytelling experience.

\pagebreak
\textbf{Visualizations and Interactions.}
To facilitate entity subsets, the \SC incorporates the data panel (Figure~\ref{fig:sc-overview} (1)) as list of entities and events. 
From there users can enrich or create new visualizations by adding entities, such as persons or objects, and their related events into them by the press of a button or drag and drop interactions.
In the current state of the prototype it is possible to utilize timelines and geo-spatial maps as visualizations within the stories.
Since only one visualization per slide is allowed, the choice of the used visualization type depends mainly on the specific focus of the slide either on temporal or geo-spatial contextualization.
Both of them are similarly designed to display entities and their related events, supported by various coloring modes to differentiate visually between entity-identities, event-kinds and a temporal color scale \cite{bach2015} ranging from begin to end of the entity's or event set's time period. 
To minimize visual clutter both visualizations contain the option to cluster events in donut or dot cluster glyphs as shown in Figure~\ref{fig:vas}.
The various interactive elements of the visualizations and the interface are linked together to react on common interactions such as mouse-overs.
Selecting events within the visualizations allows to focus on them during the story viewing to enable seamless transitions with animations throughout the story slides during the presentation in the InTaVia Story Viewer.

\textbf{Annotations and Content Chunks.}
By providing additional context and information through annotating slides with multimedia contents such as text, images, and videos users add the narration and increase tangibility. 
More advanced content types such as multiple-choice quizzes and the HTML-container hold the potential for gamification and further interactivity. 
Because of the flexible layout options, the various content types can be combined with the visualizations and arranged together.
For example multimedia quizzes are possible through the alignment of images/videos and quizzes.
The HTML ("Hypertext Markup Language") content type acts as container to include further applications such as three-dimensional object renderings, other web-applications or further visualizations, but also any other content by rendering of HTML.
Each of the content chunks is adjustable through a settings dialog to personalize the story's visual elements and create visually compelling and tailored narratives.



\smallskip

\subsection{Story Viewer}

The \SV is an integral part of the Visual Storytelling Suite, designed to enable users to preview and experience the stories created in the Story Creator. 
The Story Viewer brings the created stories with their configured visualizations and content chunks to life, providing users with an interactive and engaging narrative experience. 
It enables seamless transitions between slides, smooth visualization rendering, and includes interactive elements, fostering exploration, user engagement and immersion in the storytelling process (Figure~\ref{fig:sV}).

\section{Case Study: Traveling with Albrecht D\"urer}
\label{sec:caseStudyAlbrecht}

We illustrate the different functionalities of the Storytelling Suite with an exemplary story on the influences of Albrecht D\"urer's travel activities on his oeuvre, which was generated by the D\"urer expert Anja Grebe \cite{grebe2012,grebeNL2013}. 

Albrecht D\"urer (1471-1528) counts among the central figures of Western art history. Thanks to his extensive travel activities and his widely sold prints, his works quickly spread all over the globe and now form the pride of museums and collections worldwide \cite{grebe2013}. D\"urer is arguably also one of the best biographically documented artists from the early modern times. One of the best documented parts of his life is the so-called Journey to the Netherlands, thanks to a related travel diary and to other contemporary sources \cite{grebeNL2013}.


For Albrecht D\"urer’s life, three major journeys (two to Italy and one to the Netherlands) play an essential role, as they have been deemed an undeniable factor and driver of both the development of D\"urer’s style and the development of his transnational reputation. Based on a geographical analysis of Dürer's travel activities and related cultural objects, two stories have been generated: a macro story giving an overview on his life and work (see \url{https://youtu.be/yRzNtX7Dmow}) and a more fine-grained story focusing on his journey to the Netherlands.
While these two stories have been developed separately, they could also be presented in a nested fashion, where the user can start from the biographical macro-story first, to explore parts of his life-- such as the journey to the Netherlands--in greater detail on demand in nested slides. 





\section{Discussion}

Storytelling guidelines often assume that an intention or message stands above or behind every story that should be conveyed. However, in order to get there, a large number of practices to process and analyze different sources of information have to be conducted and orchestrated. In the case of cultural heritage topics, the information has to be found, collected and unified, before it can be processed elaborately. InTaVia demonstrates how this workflow can be supported in one integrative platform---enabled by a modular architecture which provides tools to support these different but interrelated steps of the workflow: (i) searching for data on related cultural objects and actors in an integrated knowledge graph; (ii) (re-)creating missing data aspects, (iii) inspecting and curating relevant data; (iv) visually analyzing and representing the data; and (v) creating stories with these data and visualizations---as the presented case study shows in an illustrative manner.

While many of the resulting stories created by cultural heritage experts are compelling, we realized that there are several limitations associated with our workflow: \\

(1) A decisive factor for the productive exploratory analysis of CH data is a certain richness and interconnection of the knowledge graph---which is not given across the whole reach of the existing InTaVia graph.\footnote{For some well-connected clusters within the data, we can demonstrate the feasibility of the approach, e.g. by searching for actors related to Tuusula (an artist community at a lake in Finland) or the Wiener K\"unstlerhaus (an art association in Vienna).} To overcome this problem, we take several routes: First, we currently aim to increase the interconnectedness of entities by means of different NLP and AI-based enrichment procedures. Secondly, we allow users to import their own datasets from different sources (manually generated or mapped to the InTaVia data model) and use our modules for visual analysis and storytelling on them.

(2) Even though everyone tells stories, not everyone is a skilled storyteller---or knows how to select data and design visualizations. We currently follow several threads and strategies to increase guidance for users: a) We conduct a survey on visualization-based storytelling in the digital humanities to explore and sketch out the corresponding design space. b) We study how specific design features of visualization-based stories influence the attention, interest, and engagement of recipients. c) We design guiding UI elements, which will be included in the interface to support users unfamiliar to visualization-based storytelling.


\medskip

The outlined workflow approach to visualization-based storytelling offers a chance to reach out to domain-experts in the cultural heritage field, to catalyze intra- and inter-disciplinary collaboration when creating such stories, and to inform and provide the interested public with compelling stories on cultural topics in the context of museums, in cultural tourism, but also in classrooms. By creating compelling visualization-based stories for various users and application domains, we can catch the attention of casual users and create awareness on important cultural topics in a wide range of arts and humanities fields.

\bigskip

\acknowledgments{We would like to thank the whole InTaVia team for the excellent and constructive collaboration, which enabled the presented work. The InTaVia project project has received funding from the European Union’s Horizon 2020 research and innovation programme under grant agreement No. 101004825.}

\bigskip

\bibliographystyle{abbrv-doi}

\bibliography{vis4dh}

\end{document}